\documentclass[aps,prc,twocolumn,superscriptaddress,nofootinbib]{revtex4-1}
\usepackage{amsmath,graphicx}
\usepackage{color}\usepackage{xcolor}

\begin{document}

\title{Non-Gaussian eccentricity fluctuations}

\author{Hanna Gr\"onqvist}
\author{Jean-Paul Blaizot}
\author{Jean-Yves Ollitrault}
\affiliation{
Institut de physique th\'eorique, Universit\'e Paris Saclay, CNRS, CEA, F-91191 Gif-sur-Yvette, France} 
\date{\today}

\begin{abstract}
We study the fluctuations of the anisotropy of the energy 
density profile created in a high-energy collision at the LHC.
We show that the anisotropy in harmonic $n$ has generic non-Gaussian
fluctuations. 
We argue that these non-Gaussianities have a  universal character for small systems 
such as p+Pb collisions, but not for large systems such as Pb+Pb
collisions where they depend on the underlying non-Gaussian statistics
of the initial density profile. 
We generalize expressions for the eccentricity cumulants
$\varepsilon_2\{4\}$ and $\varepsilon_3\{4\}$
previously obtained within the independent-source model to a  general fluctuating
initial density profile. 
\end{abstract}
\maketitle

\section{Introduction}

Anisotropic flow in heavy-ion collisions~\cite{Heinz:2013th} can be
simply understood as the hydrodynamic response to spatial anisotropy in
the initial state~\cite{Luzum:2013yya}. 
The largest components of anisotropic flow are elliptic flow, $v_2$,
and  triangular flow, $v_3$. In hydrodynamics, 
both are determined to a good approximation by linear 
response~\cite{Gardim:2014tya,Niemi:2012aj,Gardim:2011xv} to the 
eccentricity $\varepsilon_2$~\cite{Alver:2006wh} and
triangularity $\varepsilon_3$~\cite{Alver:2010gr,Teaney:2010vd} 
of the initial energy density profile. 
As a consequence, the probability distribution of anisotropic
flow~\cite{Aad:2013xma} directly constrains the initial 
geometry~\cite{Renk:2014jja,Yan:2014nsa}.

The fluctuations of the initial anisotropy $\varepsilon_n$ 
are to a first approximation  
Gaussian~\cite{Voloshin:2007pc,Blaizot:2014nia}. 
When the anisotropy is solely due to fluctuations (that is, with the
notable exception of $v_2$ in non-central nucleus-nucleus
collision, which is mostly driven by the eccentricity in the reaction plane), 
non-Gaussianities can be measured directly using higher-order
cumulants of the distribution of $v_n$, for instance the order 4
cumulant $v_n\{4\}$. 
Non-Gaussian flow fluctuations have first been seen through $v_3\{4\}$
in Pb-Pb collisions~\cite{ALICE:2011ab,Chatrchyan:2013kba}. 
Similar non-Gaussianities are seen in initial-state 
models of $\varepsilon_3$~\cite{Bhalerao:2011ry}.
$v_2\{4\}$ has also been measured in p-Pb 
collisions~\cite{Aad:2013fja,Chatrchyan:2013nka}, and is also
predicted by standard initial-state models~\cite{Bozek:2013uha}. 

The question therefore arises as to what non-Gaussianities can tell us about the density fluctuations in 
the initial state: do they reveal interesting features of the dynamics, or are they the result of some general constraints?  
It has been pointed out for instance that the condition $|\varepsilon_n|\le 1$
alone generates a universal non-Gaussian component~\cite{Yan:2013laa}, 
which matches recent measurements of higher-order cumulants 
$v_2\{6\}$ and $v_2\{8\}$ in p-Pb
collisions~\cite{Khachatryan:2015waa}. 
On the other hand, this is known to be only approximate. 
General analytic results about the statistics of 
$\varepsilon_n$ can be obtained within a simple model where the 
initial density profile is a superposition of pointlike, 
independent sources~\cite{Bhalerao:2006tp}. 
Non-gaussianities arise typically as corrections to the central 
limit~\cite{Floerchinger:2014fta}. 
Expressions of $\varepsilon_2\{4\}$~\cite{Alver:2008zza} and 
$\varepsilon_3\{4\}$~\cite{Bhalerao:2011bp} 
reveal a non-trivial dependence on the initial density profile, thus
breaking the universal behavior just mentioned, as will be illustrated in
Sec.~\ref{s:identical}. 

The goal of this paper is to assess more precisely what
the non-Gaussianity of anisotropy fluctuations  may tell us about
the initial density profile and its fluctuations, thereby extending the study initiated  in
Ref.~\cite{Blaizot:2014wba}. 
For simplicity, we restrict ourselves to the case of central
collisions, i.e. $b=0$, where initial anisotropies are solely due to
fluctuations. In the next section we recall general definitions of eccentricities and the cumulants of their probability distribution. Then in Sec.~\ref{s:identical}, we review known results from the independent source model. 
In Sec.~\ref{s:perturbative}, we carry out a perturbative
analysis for a general fluctuating density distribution, assuming that the fluctuations are small and uncorrelated.
In Sec.~\ref{s:mc}, these 
perturbative results are compared with full Monte Carlo
simulations in order to assess the validity of the perturbative
expansion. Conclusions are presented in Sec.~\ref{sec:discussion}. Technical material is gathered in several appendices.


\section{Initial anisotropies}
\label{s:anisotropies}

We first recall the definitions of the 
anisotropy $\varepsilon_n$ and of the cumulants 
$\varepsilon_n\{2\}$ and $\varepsilon_n\{4\}$. 
We denote by $\rho(z)$ the energy density in a given event, where 
$z=x+iy$ is the complex coordinate in the transverse plane. 
The complex Fourier anisotropies~\cite{Teaney:2010vd,Qiu:2011iv}
are defined by  
\begin{equation}
\label{defepsn}
\varepsilon_n=\frac{\int_z   z^n \rho(z)}
{\int_z   |z|^n \rho(z)},
\end{equation}
where we use the short hand $\int_z=\int {\rm
  d}x{\rm d}y$ for the integration over the transverse  plane. 
The definition (\ref{defepsn}) assumes that the center of the density
lies at the origin. In an arbitrary coordinate system, one must
replace $z$ with $z-z_0$, where 
$z_0=\int_zz\rho(z)/\int_z\rho(z)$ is the center of the distribution. 
We refer to this correction as to the ``recentering'' correction.

The density $\rho(z)$ fluctuates  event to event, which entails fluctuations of the eccentricities 
$\varepsilon_n$. There is therefore an associated probability
distribution of $|\varepsilon_n|$. Assuming that
$v_n$ is proportional to $|\varepsilon_n|$ in every event, the
probability of $|\varepsilon_n|$ is, up to a rescaling, the measured
probability distribution of $v_n$~\cite{Aad:2013xma}.
Experimental observables involve even moments of this distribution, which are conveniently combined into cumulants. The first 2
cumulants $\varepsilon_n\{2\}$ and $\varepsilon_n\{4\}$ are defined
as~\cite{Borghini:2001vi,Miller:2003kd}:
\begin{eqnarray}
\label{defcumul4}
\varepsilon_n\{2\}^2&\equiv&\langle|\varepsilon_n|^2\rangle\cr
\varepsilon_n\{4\}^4&\equiv& 2\langle |\varepsilon_n|^2\rangle^2-\langle
|\varepsilon_n|^4\rangle, 
\end{eqnarray}
where angular brackets denote averages over events in a given centrality
class. 

Note that cumulants of anisotropic flow, which are defined similarly,
with $\varepsilon_n$ replaced by $v_n$, were originally 
introduced~\cite{Borghini:2001vi}  in order to eliminate nonflow
correlations: The idea was that higher-order cumulants such as
$v_n\{4\}$ would isolate collective motion, and that the difference
between $v_n\{2\}$ and $v_n\{4\}$ was due to jets and other sources
not driven by collective flow. It was then recognized that 
nonflow correlations are largely suppressed by rapidity gaps~\cite{Adler:2003kt} and that
the difference between $v_n\{2\}$ and $v_n\{4\}$ mostly comes from
fluctuations in the initial geometry, i.e., it originates from the
difference between $\varepsilon_n\{2\}$ and $\varepsilon_n\{4\}$. 

If the anisotropy is solely due to fluctuations, and if the distribution of 
anisotropy fluctuations is Gaussian \cite{Voloshin:2007pc}, 
$\varepsilon_n\{4\}$ vanishes. Thus a non-vanishing $\varepsilon_n\{4\}$ directly
reflects the non-Gaussianity of anisotropy fluctuations.
If one assumes that $v_n$ is proportional to $\varepsilon_n$ in every
event for $n=2,3$, then 
the observation of a positive $v_2\{4\}$ in proton-nucleus
collisions~\cite{Aad:2013fja,Chatrchyan:2013nka} 
implies that $\varepsilon_2\{4\}^4>0$, 
and the observation of a 
positive $v_3\{4\}$~\cite{ALICE:2011ab,Chatrchyan:2013kba} in
nucleus-nucleus collisions, at all centralities, implies that 
$\varepsilon_3\{4\}^4>0$. 
Since the anisotropy is due to fluctuations in both cases, 
this in turn implies that these fluctuations are not Gaussian.

As already stated, our goal  is to see  what these results tell us about the fluctuations of
the density $\rho(z)$. The task is complicated by the fact that  the relation between  density and  eccentricity fluctuations is not a direct one, because the  relation (\ref{defepsn}) between $\rho(z)$ and $\varepsilon_n$ is non linear.  Also, the same relation  (\ref{defepsn}) shows that $\rho(z)>0$ ensures that $|\varepsilon_n|\le 1$. This puts a constraint on the allowed range of local density fluctuations. 
Our efforts will mostly focus on  the relations between the cumulants of the density fluctuations and those of the eccentricity fluctuations. 
Our study is limited to $\varepsilon_n\{2\}$ and
$\varepsilon_n\{4\}$, but higher-order cumulants such as
$\varepsilon_n\{6\}$ could be studied 
in a similar way.

\section{Identical sources}
\label{s:identical}

We first recall known analytical  results  obtained within a simple
model where 
the energy density is represented by a sum of identical, pointlike sources,
much as in a Monte Carlo Glauber simulation~\cite{Miller:2007ri}: 
\begin{equation}
\label{sourcemodel}
\rho(z)=\sum_{i=1}^{N}\delta(z-z_i).
\end{equation}
The positions 
$z_i$ of the sources, $i=1,\cdots,N$, are independent random variables with
probability $p(z_i)$ ($\int_zp(z)=1$)  and $N\ge 2$ is fixed. 
With this normalization, the total energy is $\int_z\rho(z)=N$. It is 
dimensionless. 
Since the sources are independent, the statistics of the fluctuations of 
$\rho(z)$ is formally equivalent to that of the  density fluctuations of a two-dimensional
ideal gas of $N$ particles with an average density profile
$\langle\rho(z)\rangle=Np(z)$. 
Inserting Eq.~(\ref{sourcemodel}) into (\ref{defepsn}), one obtains
\begin{equation}
\label{epspointlike}
\varepsilon_n=\frac{\sum_{i=1}^N (z_i-z_0)^n}{\sum_{i=1}^N |z_i-z_0|^n},
\end{equation}
where $z_0=(1/N)\sum_{i=1}^Nz_i$ is the center of the distribution. 
Throughout this paper, we assume for simplicity that
$\langle\rho(z)\rangle$ has radial symmetry and depends only on
$r\equiv |z|$, that is, we consider collisions at zero impact parameter. 
The anisotropy $\varepsilon_n$ still differs from zero in general because 
the number of sources $N$ is finite ($N$ controls the strength of the fluctuations, which vanishes as $N\to \infty$). 
The non linear dependence between $\rho(z)$ and $\varepsilon_n$ is reflected here in a non linear dependence of $\varepsilon_n$ on the position of the sources. This makes the analytical calculation of the distribution of $\varepsilon_n$ difficult for an arbitrary $p(z_i)$.

\subsection{An exact result}
\label{s:exact}

In the particular case where the average density profile is  Gaussian,  
$p(z_i)\propto\exp(-|z_i|^2/R_0^2)$,  
the probability distribution
of $|\varepsilon_2|$ can be 
calculated exactly~\cite{Yan:2013laa,Ollitrault:1992bk}:\footnote{The exact result in
  Ref.~\cite{Ollitrault:1992bk} is derived without the recentering
  correction. However, it can be shown that the recentering correction
  amounts to replacing $N$ with $N-1$.} 
\begin{equation}
P(|\varepsilon_2|)=
(N-2)|\varepsilon_2|\left(1-|\varepsilon_2|^2\right)^{\frac{N}{2}-2}.
\end{equation}
Eqs.~(\ref{defcumul4}) and  straightforward integrations then give the
first cumulants: 
\begin{eqnarray}
\label{exactcumul}
\varepsilon_2\{2\}&=&\sqrt{\frac{2}{N}}\cr
\varepsilon_2\{4\}&\equiv& \left(\frac{16}{N^2(N+2)}\right)^{1/4}.
\end{eqnarray}
In the limiting case $N=2$,
the energy consists of two pointlike
spots, therefore $|\varepsilon_2|=1$ for all events,
which implies
$\varepsilon_2\{2\}=\varepsilon_2\{4\}=1$.

\begin{figure}[h]
\begin{center}
\includegraphics[width=\linewidth]{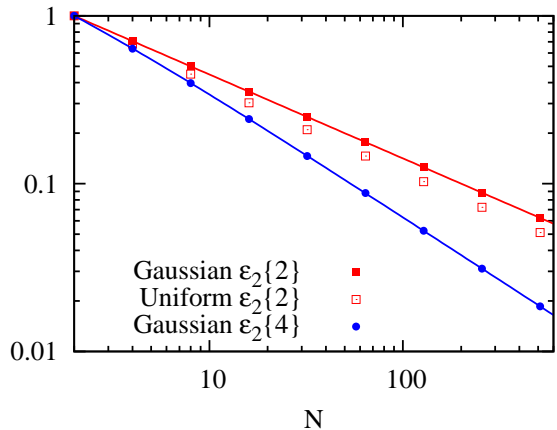} 
\end{center}
\caption{(Color online) 
Closed symbols: values of $\varepsilon_2\{2\}$ and $\varepsilon_2\{4\}$
obtained in a Monte Carlo simulation of the Gaussian
independent-source model. 
Lines: exact result given by Eq.~(\ref{exactcumul}). 
Open squares display, for sake of comparison, the values of
$\varepsilon_2\{2\}$ for a uniform average density profile (Sec.~\ref{s:asymptotic}). 
Statistical errors are smaller than symbols. 
\label{fig:exact}
}
\end{figure}    

Both $\varepsilon_2\{2\}$ and $\varepsilon_2\{4\}$ vanish in the limit
$N\to\infty$, as expected since the average density profile is
isotropic. 
$\varepsilon_2\{4\}$ decreases faster than $\varepsilon_2\{2\}$ 
because eccentricity fluctuations become more and more Gaussian in the limit of
large $N$. 
As we shall see below, 
the scaling laws $\varepsilon_2\{2\}\propto N^{-1/2}$ and 
$\varepsilon_2\{4\}\propto N^{-3/4}$ are general, in the source model, for
fluctuation-dominated eccentricities in the limit $N\gg
1$~\cite{Floerchinger:2014fta}.  

The identical source model can easily be implemented through Monte
Carlo simulations, by sampling the positions of the source $z_i$ according to the distribution $p(z)$, 
for a large number of events. 
Numerical results are shown in Fig.~\ref{fig:exact}. 
They are compatible with the exact result for all $N$, as they should. 

Eliminating $N$ between the two equations (\ref{exactcumul}), one
obtains the following relation between $\varepsilon_2\{2\}$ and 
$\varepsilon_2\{4\}$:
\begin{equation}
\label{eps4vseps2}
\varepsilon_2\{4\}=\varepsilon_2\{2\}^{3/2}\left(\frac{2}{1+\varepsilon_2\{2\}^2}\right)^{1/4}.
\end{equation}
It has been conjectured~\cite{Yan:2013laa} that this relation, which
effectively takes into account the constraint $|\varepsilon_2|<1$, holds to
a good approximation for all models of initial conditions, and also
for $\varepsilon_3$.   However, we shall see on explicit examples that this is not always the case. 

\subsection{Perturbative results}
\label{s:asymptotic}

More general results, i.e., valid for an arbitrary 
  average density profile $p(z)$ and when $N\gg
1$, have been obtained for $\varepsilon_2$~\cite{Alver:2008zza} and
$\varepsilon_3$~\cite{Bhalerao:2011bp}, 
by treating fluctuations as a small parameter, as we shall explain later. 
To leading order in $1/N$, one obtains 
\begin{eqnarray}
\label{alver}
\varepsilon_n\{2\}^2&=&\frac{1}{N}\frac{\langle r^{2n}\rangle}{\langle r^{n}\rangle^2}\cr
\varepsilon_n\{4\}^4&=&\frac{1}{N^3}\left(
-\frac{8\langle r^{2n}\rangle^3}{\langle r^n\rangle^6}
+\frac{8\langle r^{3n}\rangle\langle r^{2n}\rangle}{\langle
  r^n\rangle^5}\right.\cr
&&
-\frac{\langle r^{4n}\rangle}{\langle r^n\rangle^4}
+\left.\frac{2\langle r^{2n}\rangle^2}{\langle r^n\rangle^4}
\right),
\end{eqnarray}
where angular brackets denote average values taken with $p(z)$ 
(or, equivalently, the average density profile
$\langle\rho(z)\rangle$), and $r\equiv |z|$.  

Note that $\varepsilon_n\{4\}^4$ is the sum of two positive and two
negative terms, and there are typically large cancellations. For a
2-dimensional Gaussian   average density profile, for instance, 
$\langle r^{2k}\rangle=k!R_0^{2k}$ and Eq.~(\ref{alver}) yields for $n=2$:
\begin{eqnarray}
\label{alvergauss}
\varepsilon_2\{2\}^2&=&\frac{2}{N}\cr
\varepsilon_2\{4\}^4&=&\frac{1}{N^3}\left(-64+96-24+8\right)=\frac{16}{N^3},
\end{eqnarray}
in agreement with the exact result (\ref{exactcumul}) for $N\gg 1$.

\begin{figure}[h]
\begin{center}
\includegraphics[width=\linewidth]{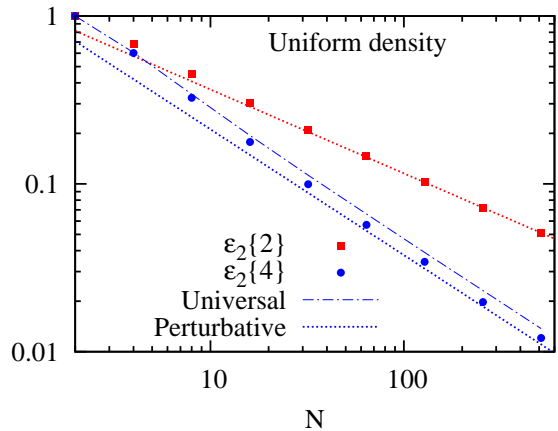} 
\end{center}
\caption{(Color online)
Same as Fig.~\ref{fig:exact} for a uniform density distribution in a
disk. 
Symbols are values obtained in a Monte Carlo simulation, 
and dotted lines are straight lines corresponding to the perturbative 
formulas (\ref{alveruniform}).  
Dash-dotted line, labeled ``Universal'': value of
$\varepsilon_2\{4\}$ derived using Eq.~(\ref{eps4vseps2}),
where $\varepsilon_2\{2\}$ in the right-hand side is taken from the
Monte Carlo result.
\label{fig:uniform}
}
\end{figure}

For a generic average density profile $\langle \rho(r)\rangle$, 
the relative magnitudes of the four terms in the expression of 
$\varepsilon_n\{4\}^4$, Eq.~(\ref{alver}), may vary. 
The Cauchy-Schwarz inequality $\langle xy\rangle^2\le\langle
x^2\rangle\langle y^2\rangle$ with $x=r^{n/2}$ and $y=r^{3n/2}$ proves
that the second term
is always larger than the first term (in absolute
 magnitude). The first term is itself at least four times larger than
 the  fourth term. But the magnitude of the third term may vary
 significantly, so that  the sign of 
$\varepsilon_n\{4\}^4$ may be negative if the density decreases slowly
 for large $r$. 
Therefore, the observation that $v_n\{4\}^4>0$ in experiments, which
implies that $\varepsilon_n\{4\}^4>0$ if $v_n$ is  proportional to
$\varepsilon_n$ in every event, provides nontrivial information.

In order to illustrate the sensitivity of $\varepsilon_n$ to
$\langle\rho(r)\rangle$, 
we carry out simulations with a uniform   average density profile,
$\langle\rho(r)\rangle\propto\theta(R_0-r)$. 
Figure~\ref{fig:exact} shows that $\varepsilon_2\{2\}$ is slightly
smaller than with a Gaussian profile. This is confirmed by 
the analytic formulas Eq.~(\ref{alver}):
The moments are given by $\langle r^{2k}\rangle=R_0^{2k}/(k+1)$ and
one has
\begin{eqnarray}
\label{alveruniform}
\varepsilon_2\{2\}^2&=&\frac{4}{3N}\cr
\varepsilon_2\{4\}^4&=&\frac{368}{135N^3}.
\end{eqnarray}
Comparison with Eq.~(\ref{alvergauss}) reveals that
$\varepsilon_2\{2\}^2$ is smaller  by a factor $\frac{2}{3}$. 
With a Gaussian density profile, it may happen that a source lies far
from the center, which typically increases the anisotropy. The uniform
distribution has no tail and therefore tends to produce rounder
systems. 
Figure~\ref{fig:uniform} shows that Monte Carlo results  
converge to the perturbative values (\ref{alveruniform})  for large $N$ as expected. 

The fact that the anisotropies depend on the   average
  density profile $p(z)$  implies that the
relation between $\varepsilon_n\{4\}$ and $\varepsilon_n\{2\}$ is 
not universal:  Eq.~(\ref{eps4vseps2}) cannot always hold, as we have already indicated. Yet, as can be seen in Fig.~\ref{fig:uniform}, 
  it accounts reasonably well for numerical results at
all $N$, and is particularly accurate for small $N$ where it gives a much better
result than the asymptotic formula
(\ref{alveruniform}).  

In Sec.~\ref{s:perturbative}, the perturbative result
Eq.~(\ref{alver}) will be generalized to an arbitrary initial density profile. 
We shall be able to assign a physical interpretation to each
of the four terms in the perturbative expansion of $\varepsilon_n\{4\}^4$, namely:
\begin{itemize}
\item{The first term arises  from the non linear relation, 
   Eq.~(\ref{defepsn}), between the eccentricity and the density $\rho(z)$. 
Due to this nonlinearity, even when $\rho(z)$ has a Gaussian
distribution, the distribution of $\varepsilon_n$ can be non-Gaussian. }
\item{The second and third term arise from the genuine non-Gaussianity of the
  distribution of $\rho(z)$, that is, they are related respectively to the
  cumulants of order three and four of the density distribution. }
\item{The fourth term is due to energy conservation, namely, the constraint 
  that the total energy $N$ should be exactly the same for all events.}
\end{itemize}

\section{Generalization to an arbitrary fluctuating density profile}
\label{s:perturbative}

We now generalize the results of Sec.~\ref{s:asymptotic} to an
arbitrary (typically continuous) density profile 
$\rho(z)$. 
We write $\rho(z)=\langle\rho(z)\rangle+\delta\rho(z)$, where
$\langle\rho(z)\rangle$ is the density averaged over events and
$\delta\rho(z)$  the fluctuation. In addition, we no longer consider 
  the total energy $E=\int_z\rho(z)$ a dimensionless quantity, as in
  the identical source model.

\subsection{Small fluctuations}
\label{s:expansion} 

Radial symmetry implies $\int_z z^n \langle\rho(z)\rangle=0$ and 
Eq.~(\ref{defepsn}) can be rewritten as 
\begin{equation}
\label{epsn}
\varepsilon_n=\frac{\int_z   z^n \delta\rho(z)}
{\int_z r^n\langle\rho(z)\rangle+\int_z   r^n \delta\rho(z)},
\end{equation}
where we have neglected the recentering correction.
We introduce the shorthand notation, for any function of $z$:
\begin{eqnarray}
\label{notation}
\delta f &\equiv&\frac{1}{\langle E\rangle}\int_z f(z)\delta\rho(z)\cr
\langle f\rangle &\equiv&\frac{1}{\langle E\rangle}\int_z f(z)\langle\rho(z)\rangle,
\end{eqnarray}
where $\langle E\rangle$ is the average total energy:
\begin{equation}
\label{defeav}
\langle E\rangle=\int_z \langle\rho(z)\rangle.
\end{equation}
With this notation, Eq.~(\ref{epsn}) can be rewritten as 
\begin{equation}
\label{epsns}
\varepsilon_n=\frac{\delta z^n}
{\langle r^n\rangle+\delta r^n}=
\frac{\delta z^n}{\langle r^n\rangle}\left(
1+\frac{\delta r^n}{\langle r^n\rangle}\right)^{-1}.
\end{equation}
 We expect $\delta z^n/\langle r^n\rangle$ and $\delta r^n/\langle
r^n\rangle$ in Eq.~(\ref{epsns}) to be small  for a large system and accordingly we treat them in a perturbative expansion. 
The size fluctuation $\delta r^n$ can be neglected to leading 
order~\cite{Blaizot:2014nia}, but must be taken
into account at 
next-to-leading order, by expanding Eq.~(\ref{epsns}) in powers of
$\delta r^n$. 
One thus obtains for the moments:
\begin{eqnarray}
\label{momentsnlo}
|\varepsilon_n|^2&\simeq&\frac{\delta z^n\delta \bar z^n}{\langle r^n\rangle^2}\left(
1-2\frac{\delta r^n}{\langle r^n\rangle}+3\frac{(\delta r^n)^2}{\langle r^n\rangle^2}+\cdots\right)\cr
|\varepsilon_n|^4&\simeq&\frac{(\delta z^n)^2(\delta \bar z^n)^2}{\langle r^n\rangle^4}\left(
1-4\frac{\delta r^n}{\langle r^n\rangle}+10\frac{(\delta r^n)^2}{\langle r^n\rangle^2}+\cdots\right),\nonumber\\
\end{eqnarray}
where $\bar z=x-iy$ denotes the complex conjugate of $z$. 
To perform the average over events, 
one is then led to evaluate averages of products of $\delta
f$'s. For  instance, 
a 2-point average is of the form: 
\begin{equation}
\label{2pointav1}
\langle\delta f\,\delta g\rangle
=\frac{1}{\langle E\rangle^2}\int_{z_1,z_2} f(z_1)g(z_2)
\langle\delta\rho(z_1)\delta\rho(z_2)\rangle.
\end{equation}
More generally, terms of order $n$ in the fluctuations involve
$n$-point functions of the density field,
$\langle\delta\rho(z_1)\cdots\delta\rho(z_n)\rangle$. 
We now derive the general form of these $n$-point functions.

\subsection{Locality and cumulants}

\begin{figure}[h]
\begin{center}
\includegraphics[width=\linewidth]{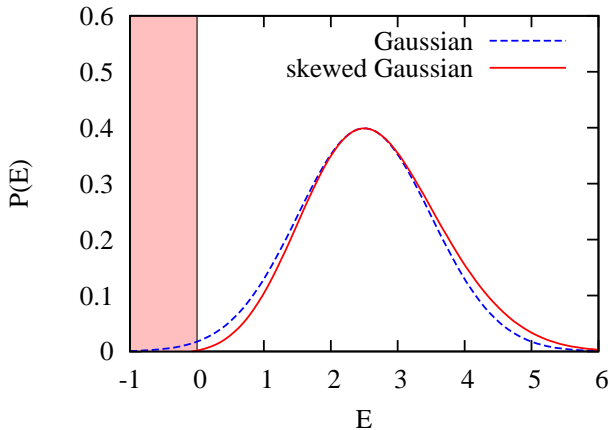} 
\end{center}
\caption{(Color online) 
Sketch of the probability of the energy $E$ in a given area. 
The Gaussian (dashed line) extends into the forbidden region $E<0$
(shaded area). Significant
improvement is obtained by skewing the Gaussian (solid line). 
\label{fig:sketch}
}
\end{figure}

We assume that that fluctuations are correlated only 
over distances much shorter than any other scale of
interest~\cite{Blaizot:2014nia}. 
A consequence of this locality hypothesis is that 
the energy $E$ contained in a transverse area $S$ much larger
than the typical area $\sigma$ of a local fluctuation has 
almost Gaussian fluctuations. 
This is seen by decomposing the area $S$ in a large number
$S/\sigma$ of independent subareas, and applying the central limit theorem. 
However, the condition   $E>0$ induces non-Gaussianities that are visible  when the relative fluctuations become sizable.
In particular, the probability is likely to have positive skew, as 
illustrated in Fig.~\ref{fig:sketch}. 
Skewness is proportional to the third cumulant of the energy
distribution. 

Consider now two areas $S_1$ and $S_2$ large compared to $\sigma$ but small compared to the total area of the system. Let $E_1$ and $E_2$ denote the energies in these two different areas. The absence of correlations between fluctuations beyond an area of size $\sigma$ implies  that $E_1$ and $E_2$ are independent variables. 
Denoting by $E=E_1+E_2$ the sum, one obtains for arbitrary $k$: 
\begin{equation}
\label{independent}
\ln\langle e^{kE}\rangle=\ln\langle e^{kE_1}\rangle+\ln\langle e^{kE_2}\rangle.
\end{equation}
This function of $k$ is called the generating function of cumulants.
The cumulant of order $n$, denoted by $\kappa_n$, is obtained by
expanding to a given order $k^n$. 
\begin{equation}
\label{defcumulants}
\ln\langle e^{kE}\rangle=\sum_{n=1}^{+\infty} \frac{k^n}{n!} \kappa_n
\end{equation}
Eq.~(\ref{independent}) shows that cumulants of the sum are sums of individual
cumulants to all orders. 

The additivity property Eq.~(\ref{independent}) implies that cumulants
of the energy in a cell scale like the transverse area $S$ of the
cell. Therefore we denote by $\kappa_n(z)\equiv \kappa_n/S$ the
density of the cumulant per unit transverse area at point $z$. At this point we note that all quantities that we are interested in are integrals of functions that are smooth on the scale of $S$. This allows us to  abandon all reference to $S$ and use a completely local formalism.
A general definition of $\kappa_n(z)$ is given in 
Appendix~\ref{s:functionalintegral} using the formalism of functional
integrals. 
The first cumulant $\kappa_1(z)$ is the average value of the energy
density, $\kappa_1(z)=\langle\rho(z)\rangle$. 
The second cumulant $\kappa_2(z)$ is the variance.  
The magnitude of fluctuations is controlled by 
  $\kappa_2(z)$. 

The identical source model of Sec.~\ref{s:identical}.
does not satisfy the locality condition
Eq.~(\ref{independent}) because the total energy $N$ is fixed by
construction, which introduces a long-range correlation. 
One recovers Eq.~(\ref{independent}) if 
$N$ is allowed to fluctuate according to a Poisson distribution, as
shown in Appendix~\ref{s:identicalsources}. 
In this case, all cumulants are equal: 
$\kappa_n(z)=\langle N\rangle p(z)$, where $\langle N\rangle$ is the
average value of $N$. 
In the general case, one can define 
an effective number of sources as follows~\cite{Blaizot:2014nia}:
\begin{equation}
\label{defneff}
N_{\rm eff}\equiv\frac{(\int_{z}  \kappa_1(z))^2}
{\int_{z} \kappa_2(z)},
\end{equation}
which coincides with $\langle N\rangle$ for identical sources.

Cumulants of order $3$ and higher vanish for a Gaussian distribution. The cumulants 
$\kappa_3(z)$ and $\kappa_4(z)$ correspond to the skewness and
kurtosis, respectively. 

\subsection{$n$-point functions}
\label{s:fieldtheory}

We show in
Appendix~\ref{s:functionalintegral} that the 2-point
function is:  
\begin{equation}
\label{2point}
\langle\delta\rho(z_1)\delta\rho(z_2)\rangle=\kappa_2(z_1)\delta(z_1-z_2), 
\end{equation}
where $\kappa_2(z)$ parametrizes the variance of the energy density at point
$z$. 
Inserting Eq.~(\ref{2point}) into Eq.~(\ref{2pointav1}), one obtains 
\begin{equation}
\label{2pointav}
\langle\delta f\,\delta g\rangle
=\frac{1}{\langle E\rangle^2}\int_{z} f(z)g(z)\kappa_2(z),
\end{equation}
where $\langle E\rangle=\int_z\kappa_1(z)$ (Eq.~(\ref{defeav})). 
The order of magnitude of the relative fluctuation is 
$
\langle\delta f\,\delta g\rangle/\langle f\rangle\langle
  g\rangle\sim 1/N_{\rm eff}$, which means that the typical order of
  magnitude of relative fluctuations $\delta f/\langle f\rangle$ in a
  given event is $1/\sqrt{N_{\rm eff}}$. 

Higher-order averages are computed in a similar way, as discussed in
Appendix~\ref{s:functionalintegral}. In particular, 
the non-Gaussian character of energy fluctuations results in 
non-trivial $3$-point averages: 
\begin{equation}
\label{3pointav}
\langle\delta f\,\delta g\,\delta h\rangle=
\frac{1}{\langle E\rangle^3}\int_{z} f(z)g(z)h(z)\kappa_3(z).
\end{equation}
This quantity is of order $1/N_{\rm eff}^2$. In a given event,
$\delta f\,\delta g\,\delta h$ is of order $1/N_{\rm eff}^{3/2}$,
but after averaging over events, the result is smaller by a factor  $1/\sqrt{N_{\rm eff}}$. 
Thus 3 and 4-point averages contribute to the same order. 
More generally, orders $2n-1$ and $2n$ both give 
contributions of order $1/N_{\rm eff}^n$. 
This implies that the expansions  of $\langle |\varepsilon_n|^2\rangle$ or $\langle |\varepsilon_n|^4\rangle$ are eventually in powers of $1/N_{\rm eff}$ rather
than $1/\sqrt{N_{\rm eff}}$.  
It also implies that 
the terms proportional to $\delta r^n$ and $(\delta
r^n)^2$ in Eq.~(\ref{momentsnlo}), 
even though they appear to be of different orders by naive
power counting, both contribute at the same (next-to-leading) order.

The fourth-order moment in Eq.~(\ref{momentsnlo}) involves terms up to order 6 in the
fluctuations. 4-point averages and higher can be reduced using Wick's 
theorem (Eqs.~(\ref{4pointav}) and (\ref{5pointav})) which breaks
them into a sum of products of lower-order terms and a connected part,
which is much smaller and vanishes for Gaussian fluctuations.

\subsection{Perturbative results}
\label{s:nlocalculation}

It is now a straightforward exercise to evaluate the moments of the
distribution of $|\varepsilon_n|$ by averaging Eq.~(\ref{momentsnlo})
over events, keeping terms up to next-to-leading order. 
We introduce the following notations:
\begin{eqnarray}
\label{aaprimebc}
a&\equiv&\frac{\langle\delta z^n\,\delta\bar z^n\rangle}{\langle
  r^n\rangle^2}\cr
a'&\equiv&\frac{\langle(\delta r^n)^2\rangle}{\langle
  r^n\rangle^2}\cr
b&\equiv&\frac{\langle\delta z^n\,\delta\bar z^n\delta
  r^n\rangle}{\langle r^n\rangle^3}\cr
c&\equiv&\frac{\langle (\delta z^n)^2\,(\delta\bar z^n)^2\rangle_c}{\langle r^n\rangle^4},
\end{eqnarray}
where the subscript $c$ in the last line denotes the connected part,
defined in Appendix~\ref{s:functionalintegral} by Eq.~(\ref{4point}). 
We have scaled by powers of $\langle r^n\rangle$ so as to obtain
dimensionless quantities. Then
$a$, $a'$, $b$ and $c$ are of order $1/N_{\rm eff}$, $1/N_{\rm eff}$,
$1/N_{\rm eff}^2$ and $1/N_{\rm eff}^3$, respectively.
Using Eqs.~(\ref{2pointav}), (\ref{3pointav}) and (\ref{rho4}), one obtains
\begin{eqnarray}
\label{aaprimebc1}
a=a'&=&\frac{\int_{z} r^{2n}\kappa_2(z)}{\left(\int_z r^n\kappa_1(z)\right)^2}
\nonumber\\
b&=&\frac{\int_z
  r^{3n}\kappa_3(z)}{\left(\int_z r^n\kappa_1(z)\right)^3}\nonumber\\
c&=&\frac{\int_z r^{4n}\kappa_4(z)}{\left(\int_z
  r^n\kappa_1(z)\right)^4}. 
\end{eqnarray}
Since one expects cumulants to be positive to all orders, these
quantities are all
positive. Both $b$ and $c$ result from the 
non-Gaussianity of density fluctuations, i.e., they are proportional to averages of the cumulants $\kappa_3(z)$ and $\kappa_4(z)$, respectively. 

The moments (\ref{momentsnlo}) can be simply expressed in terms of these
elementary building blocks using Wick's theorem (Eqs.~(\ref{4pointav})
and (\ref{5pointav})), and using radial symmetry (which implies that
$\kappa_n(z)$ only depends on $|z|$) to eliminate terms such as 
$\langle \delta z^n\,\delta
r^n\rangle$ or $\langle (\delta z^n)^2\rangle$:
\begin{eqnarray}
\label{momentsnlo2}
\langle |\varepsilon_n|^2\rangle&=&a-2 b+3 aa'\cr
\langle |\varepsilon_n|^4\rangle&=&2 a^2+c-16 a b+20 a^2 a',
\end{eqnarray}
where the first term in each line is the leading order term, and the
next terms are the next-to-leading corrections. 
The leading-order result for 
$\varepsilon_n\{2\}$ has already been obtained in
Ref.~\cite{Blaizot:2014nia}, namely:
\begin{equation}
\label{eps2lo}
\varepsilon_n\{2\}^2\equiv\langle |\varepsilon_n|^2\rangle=a
=\frac{\int_{z} r^{2n}\kappa_2(z)}{\left(\int_z r^n\kappa_1(z)\right)^2}.
\end{equation}
Terms of order $1/N_{\rm eff}^2$ cancel in the 4-cumulant 
(\ref{defcumul4}): 
\begin{equation}
\label{cumul4}
\varepsilon_n\{4\}^4=-8 a^2 a'+8 a b  -c,
\end{equation}
where we have kept all terms of order $1/N_{\rm eff}^3$, 
which is the leading non-trivial order for this quantity. 
This equation is one of the main results of this
  article. 
Together with Eq.~(\ref{aaprimebc1}), 
it expresses the non-Gaussianity of eccentricity fluctuations in terms
of the statistical properties of the underlying density field, in the regime where the perturbative expansion is valid. 

We can check the result for identical, pointlike sources where
$\kappa_n(z)=\langle N\rangle p(z)$. 
Eq.~(\ref{aaprimebc1}) then reduces to 
\begin{eqnarray}
\label{aaprimebcp}
a=a'&=&\frac{1}{\langle N\rangle}\frac{\langle r^{2n}\rangle}{\langle r^n\rangle^2}\cr
b&=&\frac{1}{\langle N\rangle^2}\frac{\langle r^{3n}\rangle}{\langle r^n\rangle^3}\cr
c&=&\frac{1}{\langle N\rangle^3}\frac{\langle r^{4n}\rangle}{\langle r^n\rangle^4}.
\end{eqnarray}
Inserting these expressions into Eq.~(\ref{cumul4}), one recovers
the first three terms in Eq.~(\ref{alver}), if one replaces $N$ with 
$\langle N\rangle$. 
The missing (fourth) term is due to energy conservation (i.e., the
condition that $N$ is fixed), which breaks locality. 
As shown in
Appendix~\ref{s:identicalsources}, the missing term appears as a
contribution to $c$.

We now discuss the  case of 
a general density $\rho(x)$. 
If density fluctuations were Gaussian, $b$ and $c$ would vanish, 
which would result in $\varepsilon_n\{4\}^4<0$. 
The only positive contribution is the second term in Eq.~(\ref{cumul4}),
which originates from the third cumulant of the density
fluctuations.\footnote{Energy conservation also gives a positive
  contribution, but it is typically much smaller} 
Therefore the observation of a positive $v_n\{4\}$ in experiments is
by itself a clear indication that the density field has positive skew. 

For simplicity, we have neglected energy conservation and the
recentering correction.
Energy conservation is important in practice because 
experimental analyses are essentially done at fixed energy:
Experiments use as a proxy for impact parameter 
an observable dubbed ``centrality'' which is
typically based on the energy deposited in a detector 
(a scintillator in the case of ALICE~\cite{Abelev:2013qoq}), which is 
strongly correlated with the total energy. 
Therefore one can essentially consider 
that the total energy $E=\int_z\rho(z)$ is fixed in a narrow centrality
class. This imposes a constraint on the density fluctuations,
which modifies the expressions of $a'$, $b$ and $c$, as discussed in
Appendix~\ref{s:saddlepoint}.  
However, the numerical effect on eccentricity cumulants often turns
out to be small, as the numerical study presented in the next section
will show.  
As for the effect of the recentering correction, it is discussed in detail in
Appendix~\ref{s:recentering}. It brings additional terms to 
Eq.~(\ref{momentsnlo2}), but these terms cancel in
$\varepsilon_n\{4\}^4$ so that Eq.~(\ref{cumul4}) is unchanged. 

\section{Monte Carlo simulations}
\label{s:mc}

In this section we present results of numerical simulations. Our goal is twofold. First,
we want to assess the domain of validity of the results 
obtained in Sec.~\ref{s:perturbative}: How large must the system be
for the expansion in powers of fluctuations  to be valid?
Second, we want to go beyond the identical source model of 
Sec.~\ref{s:identical} and test the perturbative results in a more
general situation where cumulants of the density $\kappa_n(z)$ depend
on the order $n$. This will be achieved by weighing the sources
differently so that they are no longer identical, as discussed in
Appendix~\ref{s:sourcecumulants}. 
All the Monte Carlo simulations in this section are done with a
Gaussian average density profile, as in Sec.~\ref{s:exact}. 

\begin{figure}[h]
\begin{center}
\includegraphics[width=\linewidth]{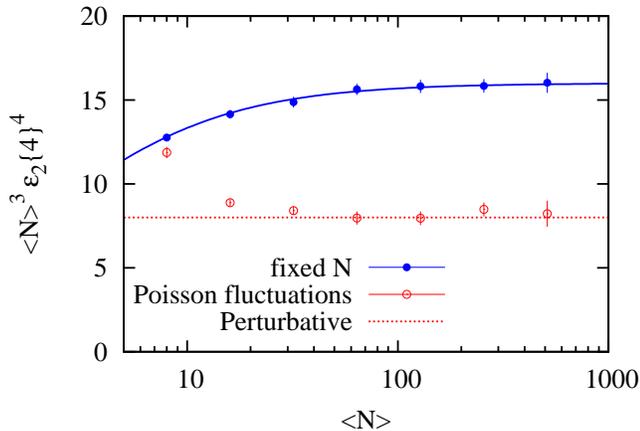} 
\end{center}
\caption{(Color online) 
$\langle N\rangle^3 \varepsilon_2\{4\}^2$ versus $\langle N\rangle$. 
Closed symbols: fixed $N$, as in Fig.~\ref{fig:exact}. 
Full line: exact result (\ref{exactcumul}). 
Open symbols: Poisson fluctuations. Dotted line: 
Perturbative result Eq.~(\ref{cumul4identical}). 
\label{fig:eps4}
}
\end{figure}

\subsection{Identical sources}
\label{s:identical2}

The first set of simulations is similar to that discussed in 
Sec.~\ref{s:exact}. 
The only difference is that the number of sources $N$ is no longer
fixed but follows a Poisson distribution,\footnote{Note that $\varepsilon_2$ is undefined for $N=0,1$, therefore we only
consider values of $\langle N\rangle$ large enough that $N=0,1$ have
probability close to 0.}
 so that we can apply Eq.~(\ref{aaprimebcp}).
Note that the Monte Carlo simulation uses a Lagrangian specification of the
density field, by sampling the position of each source, while the
derivation of Sec.~\ref{s:fieldtheory} use a Eulerian point of view,
by specifying the correlators of the density at given points. Both
Lagrangian and Eulerian descriptions are  equivalent, as shown
in Appendix~\ref{s:identicalsources}. 
 
For a Gaussian density profile, Eqs.~(\ref{cumul4}) and
(\ref{aaprimebcp}) give 
\begin{equation}
\label{cumul4identical}
\varepsilon_n\{4\}^4=
\frac{8}{\langle N\rangle^3}.
\end{equation}
Comparison with Eq.~(\ref{alvergauss}) shows that the lack of energy
conservation decreases $\varepsilon_n\{4\}^4$ by a factor 2. 
One sees in Fig.~\ref{fig:eps4} that Monte Carlo results quickly
converge to the perturbative result (\ref{cumul4identical}) for large
$\langle N\rangle$. The effect of energy conservation is 
smaller for smaller $\langle N\rangle$. 

\subsection{Negative binomial fluctuations}
\label{s:nbd}

In order to test the perturbative results of Sec.~\ref{s:perturbative}
in the more general case where cumulants differ, 
we allow for a simple generalization of Eq.~(\ref{sourcemodel}), 
by letting the energy of each source fluctuate: 
\begin{equation}
\label{wsourcemodel}
\rho(z)=\sum_{i=1}^{N}w_i\delta(z-z_i),
\end{equation}
where $w_i>0$ is the energy of source $i$, and $N$ is still distributed
according to a Poisson distribution to ensure locality. 
We assume for simplicity that the 
 fluctuations in the position ($z_i$) and the strength ($w_i$) are
  uncorrelated. 
Then $\rho(z)$ can be viewed as the density of a polydisperse ideal
gas, which generalizes the monodisperse case of
Sec.~\ref{s:identical}. 
While the cumulants of the density are all equal for identical
sources, they differ in general for weighted sources: 
$\kappa_n(z)=\langle w^n\rangle\langle N\rangle p(z)$ 
(see Appendices~\ref{s:sourcecumulants} and
\ref{s:identicalsources}). 
Inserting this expression into Eq.~(\ref{aaprimebc1}), one finds that
weights are taken into account by 
replacing $r^n$ with $wr^n$ everywhere in Eqs.~(\ref{aaprimebcp}). 

The fluctuations of the weights increase eccentricity 
fluctuations~\cite{Dumitru:2012yr,Schenke:2012fw}. 
Thus, the effective number of sources as defined by Eq.~(\ref{defneff}) is:
\begin{equation}
\label{defneff2}
N_{\rm eff}\equiv\frac{\langle w\rangle^2}{\langle w^2\rangle}\langle N\rangle.
\end{equation}
It is smaller than $\langle N\rangle$ if $w$ fluctuates.

\begin{figure}[h]
\begin{center}
\includegraphics[width=\linewidth]{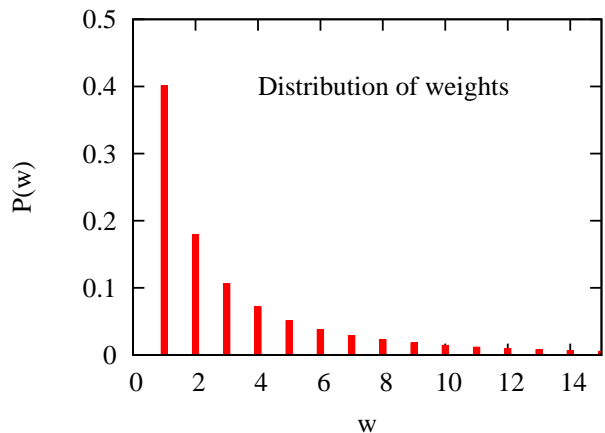} 
\end{center}
\caption{(Color online) 
Probability distribution of the energy of a single source, corresponding to
Eq.~(\ref{logarithmic2}) with $p=0.892$. 
\label{fig:weights}
}
\end{figure}    

For simplicity again, we assume that $w$ is an integer, so that it
represents a multiplicity rather than an energy. 
The probability distribution $P(w)$ used in our calculations is 
displayed in Fig.~\ref{fig:weights}. It is 
chosen in such a way that the total multiplicity 
$\sum_{i=1}^Nw_i$ follows a negative binomial distribution, in line
with observations in high-energy physics 
experiments~\cite{Giovannini:1985mz,Gelis:2009wh}. 
As shown in Appendix~\ref{s:logarithmic}, this is satisfied if 
each $w_i$ follows a logarithmic distribution: 
\begin{equation}
\label{logarithmic2}
P(w)=-\frac{1}{\ln(1-p)} \frac{p^w}{w}.
\end{equation}
This distribution depends on a single parameter $p$, which lies
between 0 and 1. 
The limit $p\to 0$ corresponds to identical sources, $w=1$. 
The larger $p$, the wider the distribution. 
Throughout this paper, we use the value $p=0.892$ corresponding to
multiplicity fluctuations at LHC energies~\cite{Kozlov:2014fqa}. 
With this distribution, Eq.~(\ref{defneff2}) gives 
\begin{equation}
\label{defneff3}
N_{\rm eff}=-\frac{p}{\ln(1-p)}
\langle N\rangle.
\end{equation}
With the chosen value of $p$, 
 $N_{\rm eff}\approx 0.4\langle N\rangle$. 
When weights are taken into account, 
Eq.~(\ref{aaprimebcp}) is replaced with (\ref{aaprimebcw}). 
These equations show that $N_{\rm eff}$ reflects incompletely the
fluctuations of the density: in particular, for a given $N_{\rm eff}$, 
the non-Gaussian contractions $b$ and $c$ increase with $p$, 
because cumulants increase rapidly with order. 
Inserting Eqs.~(\ref{aaprimebcw}) into  
Eqs.~(\ref{momentsnlo2}) and (\ref{cumul4}), one obtains for a Gaussian density profile:
\begin{eqnarray}
\label{asymptoticnbd}
\varepsilon_2\{2\}^2&=&\frac{2}{N_{\rm eff}}-\frac{12p}{N_{\rm
    eff}^2}+{\cal O}\left(\frac{1}{N_{\rm eff}^3}\right)\cr
\varepsilon_2\{4\}^4&=&\frac{8(1-3p^2)}{N_{\rm eff}^3}+{\cal O}\left(\frac{1}{N_{\rm eff}^4}\right).
\end{eqnarray}

\begin{figure}[h]
\begin{center}
\includegraphics[width=\linewidth]{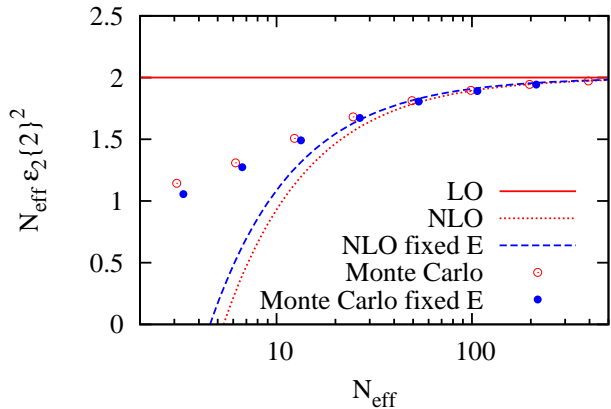} 
\end{center}
\caption{(Color online) 
$N_{\rm eff}\varepsilon_2\{2\}^2$ versus $N_{\rm eff}$. 
Symbols are results of Monte Carlo calculations, with (closed symbols)
and without (open symbols) energy conservation. 
They are slightly shifted right and left,
respectively, so that they do not overlap. 
The full line is the leading order perturbative result, 
the dotted and dashed lines are the next-to-leading
perturbative results, with (Eq.~(\ref{asymptoticnbdc})) and without 
(Eq.~(\ref{asymptoticnbd})) energy conservation. 
\label{fig:eps22}
}
\end{figure}    
Figure~\ref{fig:eps22} displays Monte Carlo results for $\varepsilon_2\{2\}$
together with the leading order and next-to-leading order perturbative
results, Eq.~(\ref{asymptoticnbd}). 
The convergence of the numerical results to the asymptotic result is
slower than for 
identical sources (see Figs.~\ref{fig:uniform} and \ref{fig:eps4}). 
The large magnitude of the next-to-leading order correction and the
fact that it overestimates the correction are signs that the
perturbative expansion diverges for small values of $N_{\rm eff}$.

\begin{figure}[h]
\begin{center}
\includegraphics[width=\linewidth]{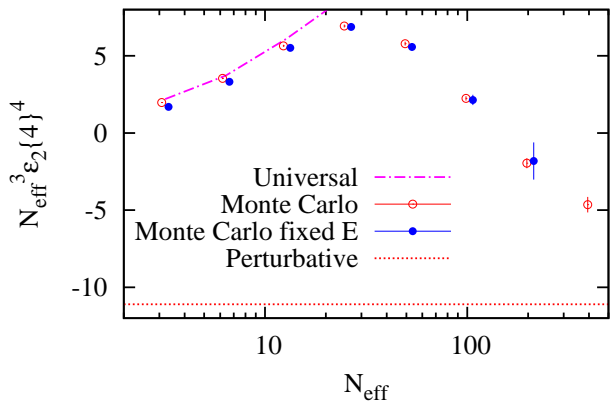} 
\end{center}
\caption{(Color online)
$N_{\rm eff}^3 \varepsilon_2\{4\}^4$ versus $N_{\rm eff}$. 
Symbols are results of Monte Carlo calculations, with (closed symbols)
and without (open symbols) energy conservation. 
As in the previous figure, they are slightly shifted to the left or to
the right. 
Dotted line: perturbative result (\ref{asymptoticnbd}). 
Dash-dotted line: Eq.~(\ref{eps4vseps2}), where $\varepsilon_2\{2\}$ in the
right-hand side is taken from the Monte Carlo simulation. 
\label{fig:eps24}
}
\end{figure}    
Figure~\ref{fig:eps24} displays our results for 
$\varepsilon_2\{4\}^4$. The perturbative result
Eq.~(\ref{asymptoticnbd}) is negative. 
On the other hand, Monte Carlo simulations return a positive
$\varepsilon_2\{4\}^4$ result for $N_{\rm eff}$ up to  $\sim 100$,
again showing that the convergence of the 
perturbative expansion is very slow. 
For values of $N_{\rm eff}$ smaller than 20, the universal scaling,
Eq.~(\ref{eps4vseps2}), gives a rather accurate result. This is not
surprising, since the condition  
$|\varepsilon_n|\le 1$ is what drives the non-Gaussianity when it is large,
which occurs if $N_{\rm eff}$ is sufficiently small. 
As $N_{\rm eff}$ increases, agreement becomes worse as, presumably,
other effects contribute to the non-Gaussianity. 
We do not have a satisfactory explanation of why the convergence to
the asymptotic value is so slow. This observation implies that the
expansion scheme chosen in Sec.~\ref{s:expansion} is not efficient
with the additional source of fluctuations considered in this section. 
Our understanding is that negative binomial fluctuations increase the
probability of a large fluctuation; and it is such a large, negative size
fluctuation $\delta r_n$ in Eq.~(\ref{epsns}) which jeopardizes the power
series expansion.

\subsection{Energy conservation}
\label{s:conservation}

Finally, we study the combined effect of negative binomial
fluctuations and energy conservation. 
In the Monte Carlo simulations, this is done by 
generating events with arbitrary energies and 
keeping only those which have the exact same 
energy\footnote{In our case, the energy is an integer. We fix it to
  the integer closest to the mean energy.} $E=\sum_{i=1}^Nw_i$.
This procedure is
time consuming, in 
particular for large systems, which is the reason why our results with
energy conservation, shown in Figs.~\ref{fig:eps22} and \ref{fig:eps24}, 
do not go as high in $N_{\rm eff}$ as results without energy
conservation. 
Numerical results show that energy conservation has a very small
effect 
for all $N_{\rm eff}$. 
This was not a priori expected: we have seen indeed that,   for identical sources,
energy conservation decreases $\varepsilon_2\{4\}^4$ by a
factor 2 (see Fig.~\ref{fig:eps4}).

Energy conservation modifies the $n$-point functions of the
density field $\rho(z)$, as shown in Appendix~\ref{s:saddlepoint}. 
As a consequence, the values of $a'$, $b$, $c$ are changed, and
Eq.~(\ref{aaprimebc1}) is replaced with (\ref{aaprimebcc}). 
With logarithmic weights, energy
conservation modifies Eq.~(\ref{aaprimebcw}) into
Eq.~(\ref{aaprimebcwc}), and the first of Eqs.~(\ref{asymptoticnbd}) is replaced
with 
\begin{equation}
\label{asymptoticnbdc}
\varepsilon_2\{2\}^2=\frac{2}{N_{\rm eff}}-\frac{8p+2}{N_{\rm
    eff}^2}+{\cal O}\left(\frac{1}{N_{\rm eff}^3}\right).
\end{equation}
The corresponding change is very modest, as can be seen in Fig.~\ref{fig:eps22}.
Furthermore, it turns out that the recentering correction, discussed
in Appendix~\ref{s:recentering}, cancels the effect of energy
conservation for a Gaussian density profile, so that the 
full next-to-leading order expression with both energy
conservation and recentering is again Eq.~(\ref{asymptoticnbd}).

We have not derived the modification of $\varepsilon_2\{4\}^4$ due to
energy conservation: it involves the connected 4-point function,
whose modifications due to energy conservation are more complicated. 
However,
the results displayed in Fig.~\ref{fig:eps24} suggest  that this
modification  may be of little relevance in practice.

\section{Discussion}
\label{sec:discussion}

We have studied the non-Gaussianity of eccentricity fluctuation by
means of Monte Carlo simulations and perturbative calculations. While
all the numerical results shown in this paper are for $\varepsilon_2$,
we have checked that conclusions also hold for the triangularity
$\varepsilon_3$.  
We have generalized the perturbative calculation of
Ref.~\cite{Alver:2008zza} to an arbitrary energy density profile,
under the sole assumption that density fluctuations at different points are uncorrelated.  

 When a perturbative expansion, which relies of the smallness of the local density fluctuations, is valid, we have obtained evidence that the non-Gaussianity of eccentricity fluctuations
largely originates from the non-Gaussianity of density
fluctuations. More specifically, the skewness and kurtosis of density
fluctuations give positive and negative contributions to
$\varepsilon_n\{4\}^4$, respectively. 
While the condition that the energy is positive
naturally generates such non-Gaussianities, their magnitude is not
universal but depends on the higher order cumulants of the density distribution.
In particular, the sign of $\varepsilon_n\{4\}^4$ is
not universal, and can be negative for a large system in the presence
of large (negative binomial) fluctuations of the multiplicity. 
Therefore, the observation that $v_3\{4\}^4$ is positive for all
centralities in Pb+Pb collisions is nontrivial.

However, Monte Carlo simulations suggest that the convergence of the 
perturbative series  can be very slow, and that results for a few hundred
sources (corresponding to the number of participants in a central
nucleus-nucleus collisions~\cite{Blaizot:2014wba}) may vary
significantly from the perturbative result. This makes it difficult to draw too definite conclusions from the present study. 
For small systems, on the other hand, we find that the universal
statistics proposed in Ref.~\cite{Yan:2013laa} is generally a good
approximation: More precisely, it works well for an effective number
of sources smaller than 15, which is the typical number of participant
nucleons for a central p+Pb collision. 
Our results thus confirm that this universal statistics should apply
to the initial anisotropies in proton-nucleus collisions. 
The observation that the values of higher order cumulants of $v_2$ 
in p+Pb collisions \cite{Khachatryan:2015waa} agree with predictions 
based on this universality therefore further supports the conclusion 
that elliptic flow in these systems originates from the initial
eccentricity $\varepsilon_2$.

\section*{Acknowledgements}
This work is 
supported by the European Research Council under the
Advanced Investigator Grant ERC-AD-267258.

\appendix

\section{Cumulants and $n$-point functions}
\label{s:functionalintegral}

In this Appendix we use functional methods to obtain 
the connected $n$-point functions and the
cumulants $\kappa_n(z)$ introduced in Sec.~\ref{s:fieldtheory}. 
The generating functional of moments is defined as: 
\begin{equation}
\label{defcumulfunctional}
{\cal Z}\left\{j(z)\right\}\equiv
\left\langle \exp\left(\int_z j(z)\rho(z)\right)\right\rangle.
\end{equation}
The $n$-point functions, such as the $2$-point function
$\langle\rho(z_1)\rho(z_2)\rangle$, are obtained by 
differentiating ${\cal Z}$ twice with respect to the auxiliary source $j(z)$ and
setting $j$ to 0. 
{\it Connected\/} $n$-point functions are obtained by differentiating 
$\ln{\cal Z}$, e.g.:
\begin{eqnarray}
\label{def2point}
\langle\rho(z_1)\rho(z_2)\rangle_c &=&
\langle\delta\rho(z_1)\delta\rho(z_2)\rangle \cr
&=&
\left.\frac{\delta^2 \ln{\cal Z}\left\{j(z)\right\}}{\delta j(z_1)\delta
  j(z_2)}\right|_{j=0}. 
\end{eqnarray}
Assuming that density fluctuations at different points are
uncorrelated entails 
that the contributions of different points to 
${\cal Z}\left\{j(z)\right\}$ factorize, 
therefore 
$\ln{\cal Z}\left\{j(z)\right\}$ can be written as an integral over $z$
of a function, which can itself be expanded in a power series 
\begin{equation}
\label{defkappaformal}
\ln{\cal Z}\left\{j(z)\right\}
=\int_z\sum_{n=1}^{+\infty} \frac{j(z)^n}{n!}\kappa_n(z). 
\end{equation}
This equation provides a formal definition of the cumulants
$\kappa_n(z)$. 
Successive differentiations of Eq.~(\ref{defkappaformal}) with respect
to $j(z)$ at $j=0$ yield connected $n$-point functions. The 2-point function is 
Eq.~(\ref{2point}). The connected 3-point and 4-point functions are
given by:
\begin{eqnarray}
\label{3point}
\langle\rho(z_1)\rho(z_2)\rho(z_3)\rangle_c&=&
\langle\delta\rho(z_1)\delta\rho(z_2)\delta\rho(z_3)\rangle\cr
&=&
\kappa_3(z_1)\delta(z_1-z_2)\delta(z_1-z_3), \nonumber\\
\end{eqnarray}
and 
\begin{eqnarray}
\label{4point}
\langle\rho(1)\rho(2)\rho(3)\rho(4)\rangle_c&\equiv&
\langle\delta\rho(1)\delta\rho(2)\delta\rho(3)\delta\rho(4)\rangle
\cr
&&-\langle\delta\rho(1)\delta\rho(2)\rangle\langle\delta\rho(3)\delta\rho(4)\rangle\cr
&&-\langle\delta\rho(1)\delta\rho(3)\rangle\langle\delta\rho(2)\delta\rho(4)\rangle\cr
&&-\langle\delta\rho(1)\delta\rho(4)\rangle\langle\delta\rho(2)\delta\rho(3)\rangle\cr
&=&\kappa_4(z_1)\delta_{12}\delta_{13}\delta_{14},
\end{eqnarray}
where $\delta\rho(i)$ stands for $\delta\rho(z_i)$, $\delta_{12}$ for
$\delta(z_1-z_2)$, etc. 
The higher-order cumulants $\kappa_3(z)$ and $\kappa_4(z)$ 
express the non-Gaussian character of the initial energy density
distribution.  

Using Eq.~(\ref{4point}), $4$-point averages can be decomposed as:
\begin{eqnarray}
\label{4pointav}
\langle\delta f\,\delta g\,\delta h\,\delta k\rangle&=&
\langle\delta f\,\delta g\rangle\langle\delta h\,\delta k\rangle\cr
&&+
\langle\delta f\,\delta h\rangle\langle\delta g\,\delta k\rangle\cr
&&+\langle\delta f\,\delta k\rangle\langle\delta g\,\delta h\rangle\cr
&&+\langle f g h k\rangle_c,
\end{eqnarray}
where the first three terms in the right-hand side, which are given by
Wick's theorem, are of order 
$1/N_{\rm eff}^2$, while the last term is a non-Gaussian correction, of
order $1/N_{\rm eff}^3$:
\begin{equation}
\label{rho4}
\langle f ghk\rangle_c=
\frac{1}{\langle E\rangle^4}\int_{z} f(z)g(z)h(z)k(z)\kappa_4(z).
\end{equation}
Higher order $n$-point functions can be expanded using Wick's theorem
in a similar way. In particular, the 5-point function gets
contributions from 2 and 3 point functions. 
\begin{eqnarray}
\label{5pointav}
\langle\delta f\,\delta g\,\delta h\,\delta k\,\delta l\rangle&=&
\langle\delta f\,\delta g\rangle\langle\delta h\,\delta k\,\delta l\rangle\cr
&&+{\rm permutations}\ (10\ {\rm terms})\cr
&&+\langle fghkl\rangle_c,
\end{eqnarray}
where the various contractions are of order $1/N_{\rm eff}^3$, 
while the connected part is of order $1/N_{\rm eff}^4$.

\section{Cumulants of local sources}
\label{s:sourcecumulants}

We  derive here the expressions of the cumulants for identical,
pointlike sources, starting with the case where each source has unit energy.
Consider the distribution of energy $E$ in an infinitesimal
transverse area $S$. This area contains a source with 
probability $\alpha\ll 1$. 
The energy $E$ in the area is $0$ with probability $1-\alpha$ and
$1$ with probability $\alpha$, therefore 
\begin{equation}
\label{source1}
\langle e^{kE}\rangle=1-\alpha+\alpha e^{k},
\end{equation}
and the generating function of cumulants is simply
\begin{equation}
\label{source2}
\ln \langle
e^{kE}\rangle=\alpha\left(e^{k}-1\right). 
\end{equation}
Using Eq.~(\ref{defcumulants}) and expanding to order $k^n$, one
obtains 
\begin{equation}
\label{sourcemoments1}
\kappa_n=\alpha.
\end{equation}
All cumulants are equal and
positive. 
If, in addition, locality is assumed, this result is generalized to an
arbitrary area by dividing it into infinitesimal areas and using the
fact that cumulants are additive. 

We assume that the numbers of sources in two separate areas are 
independent variables. 
This in turn implies that the total number of
sources $N$ follows a Poisson distribution. 

The independent source model can be generalized by allowing the energy
of each source $w$ to fluctuate with a probability 
$P(w)$. The previous results hold with the replacement of $e^k$ with $\left\langle e^{wk}\right\rangle$ in
Eqs.~(\ref{source1}) and (\ref{source2}), where brackets denote
an average taken with respect to $P(w)$. Eq.~(\ref{sourcemoments1}) is then
replaced by 
\begin{equation}
\label{sourcemoments2}
\kappa_n=\alpha\left\langle w^n\right\rangle.
\end{equation}
Cumulants are no longer equal, but still positive.

\section{Identical sources}
\label{s:identicalsources}

We derive the connected $n$-point functions of the density for the
identical source model. 
We insert $\rho(z)$ from  
Eq.~(\ref{sourcemodel}) into the generating functional
(\ref{defcumulfunctional}):
\begin{eqnarray}
\label{defcumulsource}
{\cal Z}\left\{j(z)\right\}
&=&\left\langle \prod_{i=1}^N\exp\left(j(z_i)\right)
\right\rangle\cr
&=&\sum_{N=0}^{+\infty}p_N\left(\int_z p(z)e^{j(z)}\right)^N, 
\end{eqnarray}
where $p_N$ is the probability of having $N$ sources and $p(z)$ is the
probability distribution of a source in the transverse plane. 
We now study two versions of the independent source model: 
the case where $N$ follows a Poisson distribution, and 
the case where $N$ is fixed. 

If $N$ follows a Poisson distribution, then $p_N=e^{-\langle
  N\rangle}\langle N\rangle^N/N!$. By resumming the series in 
Eq.~(\ref{defcumulsource}), one obtains
\begin{equation}
\label{functionalpoissonN}
\ln{\cal Z}\left\{j(z)\right\}
=\langle N\rangle\left(\int_zp(z)e^{j(z)}-1\right). 
\end{equation}
This equation is of the type (\ref{defkappaformal}) with
$\kappa_n(z)=\langle N\rangle p(z)$. Connected $n$-point functions are
therefore local, and cumulants $\kappa_n(z)$ are all equal, as
expected from the discussion of Appendix~\ref{s:sourcecumulants}.

For fixed $N$, Eq.~(\ref{defcumulsource}) yields
\begin{equation}
\label{functionalfixedN}
\ln{\cal Z}\left\{j(z)\right\}\equiv
N\ln\left(\int_zp(z)e^{j(z)}\right). 
\end{equation}
Thus the connected $n$-point functions to all orders are proportional
to $N$. 
Successive differentiations with respect to $j$ at $j=0$ give: 
\begin{equation}
\label{1pointsource}
\langle\rho(z)\rangle=Np(z)
\end{equation}
and, using Eq.~(\ref{def2point}):
\begin{equation}
\label{2pointsource}
\langle\delta\rho(z_1)\delta\rho(z_2)\rangle
=N\left(p(z_1)\delta(z_1-z_2)-p(z_1)p(z_2)\right).
\end{equation}
The first term in the right-hand side of Eq.~(\ref{2pointsource}) is
a local correlation.  
The second term is a disconnected term: this term results from the
constraint $\sum_z\delta\rho(z)=0$: the integral over $z$ of
Eq.~(\ref{2pointsource}) is $0$. 

The connected 3-point and 4-point functions are 
\begin{eqnarray}
\label{3pointsource}
\langle\delta\rho(z_1)\delta\rho(z_2)\delta\rho(z_3)\rangle&=&
Np(z_1)\delta(z_1-z_2)\delta(z_1-z_3)\cr
&&-Np(z_1)p(z_2)\delta(z_2-z_3)\cr
&&-Np(z_2)p(z_3)\delta(z_3-z_1)\cr
&&-Np(z_3)p(z_1)\delta(z_1-z_2)\cr
&&+2Np(z_1)p(z_2)p(z_3).
\end{eqnarray}
and
\begin{eqnarray}
\label{4pointsource}
\langle\rho(1)\rho(2)\rho(3)\rho(4)\rangle_c&=&
Np(1)\delta_{12}\delta_{13}\delta_{14}\cr
&&-N\left(p(1)\delta_{12}\delta_{13}p(4)+...\right)\cr
&&-N\left(p(1)\delta_{12}p(3)\delta_{34}+...\right)\cr
&&+2N\left(p(1)\delta_{12}p(3)p(4)+...\right)\cr
&&-6N p(1)p(2)p(3)p(4),
\end{eqnarray}
where $\delta\rho(i)$ stands for $\delta\rho(z_i)$, 
where $\delta_{12}$ stands for $\delta(z_1-z_2)$, etc., and where 
$+...$ means that one should average over all permutations,
which yield 4, 3 and 6 terms respectively for the 2nd, 3rd and 4th
lines of Eq.~(\ref{4pointsource}). 
The integral over $z_1$ is zero because of energy conservation. 

After 
inserting Eqs.~(\ref{2pointsource}), (\ref{3pointsource}) and
(\ref{4pointsource}) into the definitions of $a$, $a'$, $b$ and $c$,
Eq.~(\ref{aaprimebc}), one finds that  the condition that $N$ is fixed modifies
Eq.~(\ref{aaprimebcp}) into 
\begin{eqnarray}
\label{aaprimebcpc}
a&=&\frac{1}{\langle N\rangle}\frac{\langle r^{2n}\rangle}{\langle r^n\rangle^2}\cr
a'&=&\frac{1}{\langle N\rangle}\left(\frac{\langle r^{2n}\rangle}{\langle r^n\rangle^2}-1\right)\cr
b&=&\frac{1}{\langle N\rangle^2}\left(
\frac{\langle  r^{3n}\rangle}{\langle r^n\rangle^3}
-\frac{\langle  r^{2n}\rangle}{\langle r^n\rangle^2}\right)
\cr
c&=&\frac{1}{\langle N\rangle^3}
\left(\frac{\langle r^{4n}\rangle-2\langle r^{2n}\rangle^2}{\langle r^n\rangle^4}\right).
\end{eqnarray}
The first coefficient $a$ is unchanged. The modifications of $a'$ and
$b$ cancel in the combination $-aa'+b$ entering the expression of
$\varepsilon_n\{4\}$, Eq.~(\ref{cumul4}). 
The modification of $c$, which produces the last term in Eq.~(\ref{alver}), is due to the 3rd line of
Eq.~(\ref{4pointsource}).

\section{Energy conservation}
\label{s:saddlepoint}

We now derive the expressions of connected $n$-point functions
for an arbitrary density profile $\rho(z)$ when the total energy is fixed. 
Similar results have been obtained for momentum 
conservation~\cite{Borghini:2000cm,Borghini:2003ur}.
We denote by ${\cal Z}_E\left\{j(z)\right\}$ the generating function
corresponding to a fixed energy $E$: 
\begin{equation}
\label{diracconstraint}
{\cal Z}_E\left\{j(z)\right\}\equiv
\frac{\left\langle \exp\left(\int_z
j(z)\rho(z)\right)\delta\left(E-\int_z\rho(z)\right)\right\rangle}
{\left\langle\delta\left(E-\int_z\rho(z)\right)\right\rangle}.
\end{equation}
It is normalized so that ${\cal Z}_E\left\{j=0\right\}=1$. 
Using the integral representation of the Dirac distribution,
$\delta(x)=\frac{1}{2\pi}\int e^{-ikx} dk$, one can express 
${\cal Z}_E$ in terms of 
${\cal Z}$ using Eq.~(\ref{defcumulfunctional}): 
\begin{equation}
\label{ZE}
{\cal Z}_E\left\{j(z)\right\}=
\frac{\int_{-\infty}^{+\infty}\exp\left(-ikE\right){\cal
    Z}\{j(z)+ik\}dk}
{\int_{-\infty}^{+\infty}\exp\left(-ikE\right){\cal
    Z}\{ik\}dk}.
\end{equation}
The integral over $k$ in the numerator is evaluated using the saddle
point method. The saddle point $k_0$ is obtained by truncating 
the cumulant expansion Eq.~(\ref{defkappaformal}) to order  2, 
inserting it into (\ref{ZE}), and differentiating the exponent with
respect  to $k$: 
\begin{equation}
\label{saddlepoint}
ik_0=\frac{E-\int_z\kappa_1(z)-\int_zj(z)\kappa_2(z)}{\int_z\kappa_2(z)}.
\end{equation}
In the saddle point approximation, the integral in the numerator of
Eq.~(\ref{ZE}) is obtained by evaluating the integrand at $k=k_0$. One
thus obtains
\begin{eqnarray}
\ln{\cal
  Z}_E\left\{j(z)\right\}&=&-ik_0E+\int_z(j(z)+ik_0)\kappa_1(z)\cr
&&+\frac{1}{2}\int_z(j(z)+ik_0)^2\kappa_2(z),
\end{eqnarray}
where we have left out the contribution of the denominator which is
independent of $j$. 
The connected $n$-point functions are then obtained by differentiating 
$\ln{\cal Z}_E\left\{j(z)\right\}$ at $j=0$. 
The one-point function is the average value of the density:
\begin{equation}
\label{1pointc}
\langle\rho(z_1)\rangle=\kappa_1(z_1)+\frac{E-\int_z\kappa_1(z)}{\int_z\kappa_2(z)}\kappa_2(z_1).
\end{equation}
The first term in the right-hand side of Eq.~(\ref{1pointc}) is the
average value of 
$\rho(z)$ in the absence of energy conservation, while the second term
is the correction due to energy conservation. This equation means that
the energy excess $E-\int_z\kappa_1(z)$ is distributed in the
transverse plane proportionally to the variance $\kappa_2(z)$.

The 2-point function (\ref{2point}) becomes: 
\begin{equation}
\label{2pointc}
\langle\delta\rho(z_1)\delta\rho(z_2)\rangle=\kappa_2(z_1)\delta(z_1-z_2)- 
\frac{\kappa_2(z_1)\kappa_2(z_2)}{\int_z\kappa_2(z)}. 
\end{equation}
Note that it does not involve the value of $E$. 
One can check that the right-hand side vanishes upon integration over
$z_1$, as expected from the condition of energy conservation which
implies $\int_z\delta\rho(z)=0$. 

The connected $3$-point function is obtained by expanding 
the generating functional ${\cal Z}_E$ to first order in the
higher-order cumulants $\kappa_3$. The only effect of
energy conservation is to replace $j(z)$ by $j(z)+ik_0$, or (since
the constant term drops upon differentiation) through the substitution
\begin{equation}
j(z)\rightarrow j(z)-\frac{\int_{z'}j(z')\kappa_2(z')}{\int_{z'}\kappa_2(z')}.
\end{equation}
Eq.~(\ref{3point}) becomes: 
\begin{eqnarray}
\label{3pointc}
\langle\delta\rho(z_1)\delta\rho(z_2)\delta\rho(z_3)\rangle&=&\kappa_3(z_1)\delta(z_1-z_2)\delta(z_1-z_3)\cr 
&&-\frac{\kappa_3(z_1)\delta(z_1-z_2)\kappa_2(z_3)+{\rm
  perm.}}{\int_z\kappa_2(z)}\cr 
&&+\frac{\kappa_3(z_1)\kappa_2(z_2)\kappa_2(z_3)+{\rm
  perm.}}{\left(\int_z\kappa_2(z)\right)^2}\cr
&&-\frac{\int\kappa_3}{\left(\int\kappa_2\right)^3}\kappa_2(z_1)\kappa_2(z_2)\kappa_2(z_3)
\end{eqnarray}
The second and third lines must be summed over
circular permutations of $z_1,z_2,z_3$.
One may again check that the right-hand side vanishes upon integration
over $z_1$. 
Note that the three-point function involves cumulants of two different
orders $\kappa_2$ and $\kappa_3$, and that it is 
linear in $\kappa_3$. This linearity is due to the fact that we
evaluate the non-Gaussanity to first (linear) order. 
In the case of identical sources, where $\kappa_3(z)=\kappa_2(z)$, one
recovers Eq.~(\ref{3pointsource}). 
Note that the 2-point function and 3-point functions do not involve
$E$. The correlations are not changed by triggering on the tail of the
distribution, that is, on ultracentral collisions. 

Energy conservation modifies Eqs.~(\ref{aaprimebc}) to 
\begin{eqnarray}
\label{aaprimebcc}
a&=&\frac{\int_{z} r^{2n}\kappa_2(z)}{\left(\int_z r^n\kappa_1(z)\right)^2}\cr
a'&=&\frac{\int_{z} r^{2n}\kappa_2(z)}{\left(\int_z r^n\kappa_1(z)\right)^2}-
\frac{\left(\int_{z} r^{n}\kappa_2(z)\right)^2}{\left(\int_z r^n\kappa_1(z)\right)^2\int_{z} \kappa_2(z)}\cr
b&=&\frac{\int_z  r^{3n}\kappa_3(z)}{\left(\int_z r^n\kappa_1(z)\right)^3}-
\frac{\int_z  r^{2n}\kappa_3(z)\int_z  r^{n}\kappa_2(z)}
{\left(\int_z r^n\kappa_1(z)\right)^3\int_z  \kappa_2(z)},
\end{eqnarray} 
Note that the expression of $a$ is unchanged, so that the leading
order anisotropy (\ref{eps2lo}) is not affected by energy conservation. 
For identical sources, where $\kappa_n(z)$ is independent
  of the order $n$, Eq.~(\ref{aaprimebcc}) reduces to
  Eq.~(\ref{aaprimebcpc}).  
We have not derived the modified expression of $c$, 
which involves the connected 4-point function.

\section{Recentering correction}
\label{s:recentering}

We discuss here the effect of the recentering correction on
perturbative results. 
Our discussion is limited to $\varepsilon_2$ for simplicity. The recentering correction arises from the requirement that  the coordinate system be centered in every event.
When one takes it into account,
Eq.~(\ref{epsns}) is replaced by~\cite{Bhalerao:2011bp}
\begin{equation}
\label{eps2c}
\varepsilon_2=\frac{\delta z^2-(\delta z)^2}{\langle r^2\rangle}
\left(
1+\frac{\delta r^2}{\langle r^2\rangle}-\frac{\delta z\,\delta\bar z}{\langle r^2\rangle}\right)^{-1}.
\end{equation}
In order to evaluate the moment $\langle |\varepsilon_2|^2\rangle$ to
next-to-leading order in the fluctuations, as in
Sec.~\ref{s:nlocalculation}, we must keep all terms of order 3 and 4
in the fluctuations. The recentering correction introduces new
nontrivial contractions, in addition to those defined in
Eq.~(\ref{aaprimebc}): 
\begin{eqnarray}
\label{def}
d&\equiv&\frac{\langle(\delta z)^2(\delta\bar z)^2\rangle}{\langle
  r^2\rangle^2}=\frac{2\left(\int_{z} r^{2}\kappa_2(z)\right)^2}{\langle E\rangle^2\left(\int_z
  r^2\kappa_1(z)\right)^2}\cr
e&\equiv&\frac{\langle\delta z^2\delta\bar z^2\delta z\delta\bar z\rangle}{\langle
  r^2\rangle^3}=\frac{\int_{z} r^{4}\kappa_2(z)\int_{z} r^{2}\kappa_2(z)}{\langle E\rangle\left(\int_z
  r^2\kappa_1(z)\right)^3}
\cr
f&\equiv&\frac{\langle\delta z^2(\delta\bar z)^2\rangle}{\langle
  r^2\rangle^2}=\frac{\int_{z} r^{4}\kappa_3(z)}{\langle E\rangle\left(\int_z
  r^2\kappa_1(z)\right)^2},
\end{eqnarray}
which are all of order $1/N_{\rm eff}^2$ 
 according to the power counting of
Sec.~\ref{s:fieldtheory}. 
The first of Eqs.~(\ref{momentsnlo2}) becomes
\begin{equation}
\label{momentsnlo2rec}
\langle |\varepsilon_2|^2\rangle=a-2 b+3 aa'+d+2 e-2 f.
\end{equation}
Let us check the validity of these additional terms in the particular
case of identical sources. 
For identical sources, all cumulants are equal and the energy $E$ is
just the number of sources $N$, therefore $e=f$ and $d=2/\langle
N\rangle^2$. For a Gaussian density profile, the exact results are 
$\langle |\varepsilon_2|^2\rangle=2/N$ with recentering and 
$\langle |\varepsilon_2|^2\rangle=2/(N+1)$ without recentering. The
recentering correction is therefore $2/N-2/(N+1)\simeq 2/N^2$, in
agreement with our perturbative estimate for large $N$. 

We have checked that the recentering correction does not affect the
result  (\ref{cumul4}), i.e., it does contribute to 
$\varepsilon_2\{4\}^4$ to order $1/N^3$. 

\section{Logarithmic weights}
\label{s:logarithmic}
In this section, we explain how to choose the weights, in an
independent-source model, in such a way that the distribution of the
total energy is a negative binomial.
The negative binomial distribution is
\begin{equation}
\label{NBD}
P_{NBD}(w)=\begin{pmatrix}w+k-1\cr w\end{pmatrix}
p^w(1-p)^k,
\end{equation} 
where $p$ and $k$ are two parameters, with $0\le p< 1$ and $k>0$. If
two variables $w_1$ and $w_2$ 
are both distributed according to $P_{NBD}(w)$, then the sum $w_1+w_2$ also
follows a negative binomial distribution with the same $p$ and $k\to 2k$. 
Thus $p$ is an intensive quantity and $k$ an extensive quantity. 
In  limit of a small area, defined by $k\to 0$, 
Eq.~(\ref{NBD}) reduces to 
\begin{eqnarray}
\label{smallvolumeNBD}
P_{NBD}(0)&=&1+k\ln(1-p)\cr
P_{NBD}(w)&=&k\frac{p^w}{w}\ {\rm for}\ w\ge 1. 
\end{eqnarray}
The probability of finding a source within the small area is defined as 
\begin{equation}
\alpha\equiv 1-P_{NBD}(0)=\sum_{w=1}^{+\infty} P_{NBD}(w)= -k\ln(1-p). 
\end{equation} 
This equation gives $k$ as a function of $\alpha$. Inserting into the
second line of Eq.~(\ref{smallvolumeNBD}),
One obtains $P_{NBD}(w)=\alpha P(w)$ for $w\ge 1$,
where $P(w)$ is the distribution of $w$ for a single source, 
given by Eq.~(\ref{logarithmic2}). 

The moments of the distribution $P(w)$ can be calculated
analytically. One obtains
\begin{eqnarray}
\label{wmoments}
\frac{\langle w^2\rangle}{\langle w\rangle^2}&=&\frac{-\ln(1-p)}{p}\cr
\frac{\langle w^3\rangle}{\langle w\rangle^3}&=&\left(\frac{-\ln(1-p)}{p}\right)^2(1+p)\cr
\frac{\langle w^4\rangle}{\langle w\rangle^4}&=&\left(\frac{-\ln(1-p)}{p}\right)^3(1+4p+p^2).
\end{eqnarray}
The cumulants of the energy density are proportional to the moments of
$w$ as shown in Appendix~\ref{s:sourcecumulants}: $\kappa_n(z)=\langle
N\rangle \langle w^n\rangle p(z)$. Inserting the above expressions
into Eq.~(\ref{aaprimebc1}), one obtains
\begin{eqnarray}
\label{aaprimebcw}
a=a'&=&\frac{1}{N_{\rm eff}}\frac{\langle r^{2n}\rangle}{\langle r^n\rangle^2}\cr
b&=&\frac{(1+p)}{N_{\rm eff}^2}\frac{\langle r^{3n}\rangle}{\langle r^n\rangle^3}\cr
c&=&\frac{(1 + 4 p + p^2)}{N_{\rm eff}^3}\frac{\langle
  r^{4n}\rangle}{\langle r^n\rangle^4}, 
\end{eqnarray}
where $N_{\rm eff}$ is defined by Eq.~(\ref{defneff3}).
Eqs.~(\ref{aaprimebcw}) reduces to Eqs.~(\ref{aaprimebcp}) in the limit 
$p\to 0$, as expected. 
Inserting Eq.~(\ref{aaprimebcw}) into Eqs.~(\ref{momentsnlo2}) and
(\ref{cumul4}), one obtains Eq.~(\ref{asymptoticnbd}). 

With energy conservation taken into account, one uses
Eq.~(\ref{aaprimebcc}) instead of Eq.~(\ref{aaprimebc1}): 
\begin{eqnarray}
\label{aaprimebcwc}
a&=&\frac{1}{N_{\rm eff}}\frac{\langle r^{2n}\rangle}{\langle r^n\rangle^2}\cr
a'&=&\frac{1}{N_{\rm eff}}\left(\frac{\langle r^{2n}\rangle}{\langle r^n\rangle^2}-1\right)\cr
b&=&\frac{(1+p)}{N_{\rm eff}^2}\left(\frac{\langle
  r^{3n}\rangle}{\langle r^n\rangle^3}-\frac{\langle
  r^{2n}\rangle}{\langle r^n\rangle^2}\right). 
\end{eqnarray}

We finally consider the effect of the recentering correction. With logarithmic
weights, Eqs.~(\ref{def}) become
\begin{eqnarray}
\label{defw}
d&=&\frac{2}{N_{\rm eff}^2}\cr
e&=&\frac{1}{N_{\rm eff}^2}\frac{\langle r^{4}\rangle}{\langle r^2\rangle^2}\cr
f&=&\frac{1+p}{N_{\rm eff}^2}\frac{\langle r^{4}\rangle}{\langle
  r^2\rangle^2}.
\end{eqnarray}
Inserting Eqs.~(\ref{defw}) and (\ref{aaprimebcwc}) into 
Eqs.~(\ref{momentsnlo2rec}), one finds that the contribution to
$\varepsilon_2\{2\}^2$ 
from recentering exactly cancels the contribution from energy
conservation for a Gaussian density profile.

\end{document}